\documentstyle[twocolumn,floats,aps,epsf,prb]{revtex}
\begin{document}
\draft
\title{Effect of Interband Transitions on $c$-axis Penetration
Depth of Layered Superconductors}
\author{W. A. Atkinson\cite{bill} and J. P. Carbotte}
\address{Department of Physics and Astronomy, McMaster University,
Hamilton, Ontario, Canada L8S 4M1}
\date{\today}

\maketitle
\begin{abstract}
The electromagnetic response of a system with two planes per
unit  cell  involves,  in  addition to the usual intraband
contribution, an added interband term. These transitions affect
the  temperature  dependence and the magnitude of the zero
temperature $c$-axis penetration depth ($\lambda_c$). When the
interplane  hopping $t_\perp$  is sufficiently small, the interband
transitions dominate the low temperature behaviour of $\lambda(T)$ which
then does not reflect the linear temperature dependence of the
intraband term and in comparison becomes quite flat even for a
$d$-wave gap.  It is in this regime that the pseudogap was found in
our previous normal state calculations of the $c$ axis conductivity,
and the effects are connected.
\end{abstract}
%

\narrowtext
\section{Introduction}
Studies of the in plane low temperature penetration depth of
single crystals of twinned and untwinned YBa$_2$Cu$_3$O$_x$ (YBCO$_x$)
at optimum doping ($x=6.93$) have revealed a linear temperature 
dependence.\cite{a,b,c,d,e}
This has been confirmed in more recent measurements on single
crystals by Mao et al.\cite{f} and on thin films by de Vaulchier et
al.\cite{g} although reports of  other  power  laws  persist.\cite{h}
Extensions to underdoped samples (YBCO$_{6.6}$)
as well as to the overdoped case YBCO$_{6.99}$ confirm that this linear 
law does not depend on oxygen doping although the value of its slope is
affected. In untwinned single crystals, it has also been found
that the zero temperature value of the in-plane penetration depth
is highly anisotropic: $\lambda_b =1600$ \AA while $\lambda_a=1030$\AA where  
$b$ denotes the chain direction, and the sample doping is near optimal
($x=6.95$  with $T_c =93.2$ K).
This large anisotropy in $\lambda_{ab}(T=0)$ is, 
however, not reflected strongly in the
temperature  dependence  of the normalized penetration depth
$\lambda_i(0)/\lambda(T)$ which is found to be nearly the same for $i=a$ and
$b$,\cite{d,i,j} with that along the chain showing a slightly steeper
slope.

A linear temperature behaviour is, of course, what is
expected for a superconductor with a gap which has $d$-wave symmetry
with zeros crossing the Fermi surface. Also, the addition of Zn
impurities,\cite{e}  which are believed to act as unitary scatters, is
found to alter the temperature dependence of $\lambda_{ab}$
from linear to $T^2$, which is expected for $d$-wave.  
There are now many other
experiments that confirm the $d$-wave symmetry of the gap. Examples
are angular resolved photoemission\cite{m,n,o} (ARPES) as well as
SQUID \cite{p,q,r,s} experiments. These latter experiments have been
devised specifically to measure the phase of the gap rather than
simply probe for zeros crossing the Fermi surface as does ARPES
and penetration depth.

Detailed theoretical studies of the $a$-$b$ plane anisotropy have
been carried out by the present authors \cite{t,u,v} in a two plane
model, coupled through a transverse tunnelling matrix element
($t_\perp$).  The general conclusion of these studies is that there is a
significant amount of condensate on the chains and that the gap on
the   CuO   chains   must   be   of   the   same
order of magnitude as on the CuO$_2$ planes. The condensate on the
chains is, of course, the ultimate source of the orthorhombicity
which is reflected in the observed large in plane anisotropy for
$\lambda_a$ and $\lambda_b$ at T=0. The recent
theoretical work of O'Donovan and Carbotte\cite{aa} and of Xiang and
Wheatley\cite{bb} further  confirm  these  general  conclusions.
CITS \cite{cc,dd} experiments in which the gap is measured separately on
the chains and on the plane give added evidence that both gap
values are of roughly of the same size.
An anisotropy of similar magnitude is also seen in the DC
resistivity \cite{w,x,y,z} and optical conductivity\cite{l}. 

Recently, Hardy et al.,\cite{i} in addition to summarizing their
result for $a$-$b$ plane anisotropy in the underdoped, optimally doped
and overdoped cases, have also presented $c$-axis data.\cite{i,j,k}
For YBCO$_{6.6}$ ($T_c =59$ K), they estimate
$\lambda_c/\lambda_a \sim 31$. For YBCO$_{6.95}$ ($T_c =93.2$ K) 
$\lambda_c/\lambda_a \sim 7$ which is about
the same as for the overdoped case with x=6.99 (T =89K). They
give detailed results for the temperature variation of the ratio
$\lambda_c(0)/\lambda_c(T)$ and find, in all cases, that at low $T$, 
the curve for
the $c$-axis falls well above the linear dependence found for $\lambda_a$ and
$\lambda_b$ and may vary as a higher than linear power. A striking fact is
that the underdoped and overdoped case are found to have nearly
the  same dependence on reduced temperature $T/T_c$ while the
optimally doped case falls slightly below. A possible explanation
for the rather flat behaviour observed in the underdoped case is
that it is representative of incoherent coupling between the
chains and planes.  This view is consistent with decreasing
coupling between planar elements as oxygen is removed from the
sample.

For fully oxygenated YBCO$_x$, the experimental results on
the $c$-axis DC resistivity now appear to converge. It is found
that $\rho(T)$ is metallic in the $c$-direction and has a nearly linear
temperature dependence.\cite{w,y,ee}
Further, the anisotropy ratio $\rho_c/\rho_{ab}$ is of order 30-70 which is
consistent with the value of $\lambda_c/\lambda_a$ quoted above if we 
assume that an effective mass model is qualitatively valid. On the other
hand, when the sample is deoxygenated, the $c$-axis behaviour can
change radically with the DC resistivity $\rho_c(T)$ showing a
semiconducting-like  behaviour at low temperature ($d\rho_c/dT\,<\,0$).

Also, the ratio $\rho_c/\rho_{ab}$ is much bigger\cite{ff,gg,hh}
in the underdoped samples than for optimal
doping.  For example, at $T=100$ K, $\rho_c/\rho_{ab}$ \cite{hh} 
is of order $10^3$. This 
large value is in accordance with the value found for the
corresponding ratio of $(\lambda_c/\lambda_b)^2$. It is to be noted that 
as the temperature is increased, metallic behaviour is recovered even for
YBCO$_{6.6}$. The very large value of $\rho_c$ observed in the
underdoped case and its semiconducting behaviour has been
taken as evidence that the $c$-axis transport can not be described by
conventional three dimensional coherent Bloch transport and that it is
dominated by a different (incoherent) mechanism due to the highly
two dimensional nature of the electronic states. Evidence for
this is provided by estimates of the $c$-axis mean free path which
find that it is of the order or less than the $c$-axis spacing.

Various possible mechanisms\cite{ii,jj,kk,ll} for $c$-axis transport have
been reviewed by Zha, Cooper and Pines.\cite{ii}   In particular, Rojo
and Levin \cite{ll} have considered explicitly static and dynamic off
diagonal disorders as well as the effect of the interplane matrix
element $t_\perp$ and also include several limiting cases.  In
particular, ordinary transport and the pure incoherent case with
$t_\perp =0$ and static impurity scattering are treated. This latter case
has been discussed in detail in the work of Graf, Rainer and
Sauls.\cite{jj} It leads to a $c$-axis resistivity that is inversely
related to its in-plane value. The source of incoherent
scattering in such models is, however, not clear.  Another
interesting  idea  is the dynamic thermal dephasing process
suggested by Leggett \cite{mm} in which in-plane thermal fluctuations
can be larger than $t_\perp$ and so destroy coherent $c$-axis transport. 
This mechanism, however, cannot explain the observation that
metallic transport is restored at high temperature for the
underdoped samples.  A further model based on resonant tunneling
through a barrier layer which due to Abrikosov,\cite{abrikosov}
models the $c$-axis d.c. resistivity with some succes.
Yet another approach assumes that the copper
oxide planes cannot be described by Fermi liquid theory and spin
and charge degrees of freedom are separated.\cite{anderson}
No consensus has yet emerged about $c$-axis coupling. 

In this paper, we will be mainly interested in the $c$-axis
penetration depth. In this case,
Radtke, Kostur and Levin \cite{pp} have already given results for the
$T=0$ value and temperature dependence of $\lambda_c(T)$ in a model which
contains three distinct contributions dependent on $t_\perp$ , a pure
incoherent impurity assisted
contribution as in the work of Graf et al.\cite{jj}  and a phonon
assisted interlayer inelastic part. They find that when the
direct contribution dominates, $\lambda_c$ and $\lambda_{ab}$ will 
have the same
linear temperature dependence at low $T$ while for impurity assisted
hopping $\lambda^2_c(T)$ will show a much flatter, more $s$-wave low
temperature behaviour as is observed in experiments. There are
two problems with this scenario as applied to underdoped YBCO.
First, while $t_\perp$ may be small, the coherent part may
not be negligible. Estimates of $t_\perp$ provided by Zha et al.\cite{ii}
give $t_\perp\sim 3.0$ meV for YBCO$_{6.68}$. This value is reduced by an
order of magnitude from $t_\perp \sim $30-40 meV estimated in optimally
doped YBCO but is still significant and is certainly much
larger than the value of 0.1 meV estimated for Bi$_2$Sr$_2$CaCu$_2$O$_8$.
Secondly, the impurity assisted contribution vanishes by symmetry
for a $d$-wave superconductor if the impurity potential is isotropic
in momentum space, i.e. a $\delta$-function in direct space.  This may
not be a serious obstacle, particularly in YBCO which is
orthorhombic and so cannot have a purely $d$-wave order parameter.
A more serious problem is the microscopic origin of
this impurity scattering. This is particularly problematic since
many experiments would lead us to believe that these crystals are
in the clean limit as far as $ab$ plane properties are concerned.

   In this paper, we study a different mechanism for the $c$-axis
penetration depth. We consider the possibility that it is a
manifestation of the interband transitions which become more
important in the $c$-axis conductivity as $t_\perp$  is reduced  in
magnitude.   In  previous  work on the normal state A.C.
conductivity,\cite{qq} we found that the usual Drude-like  intraband
contribution is of order $t_\perp^0$ and the interband contribution goes
like $t_\perp^2$ for the in-plane conductivity. On the other hand,
for the $c$-axis conductivity, the intraband contribution goes like
$t_\perp^4$ while the interband remains of order $t_\perp^2$.  
It is clear then
that,  for  sufficiently small values of $t_\perp$ , the interband
contribution to the penetration depth will dominate, and $\lambda_c(T)$
may be significantly different from $\lambda_{ab}$.
It is precisely for this case that 
we found a pseudo gap in the normal state $c$ axis conductivity.  
Thus the apperence
of a pseudogap and a flat low temperature dependence in the c-axis
penetration depth are connected in our theory.

   In section \ref{two}, we present the necessary formalism. Numerical
results are given and discussed in section \ref{three}.  A brief
conclusion is included in section \ref{four}.

\section{Derivations}
\label{two}

     To  remain  simple,  we  consider two isolated planes coupled
with a transverse matrix element $t_\perp$.  The  uncoupled  planes  have
dispersion
\begin{equation}
\xi_1 = - 2\sigma_1 [\cos(k_x)+\cos(k_y) - 2B \cos(k_x)\cos(k_y)]-\mu_1,
\label{1}
\end{equation}
representing the tetragonal CuO$_2$ plane and
\begin{equation}
\xi_2 = - 2 \sigma_2 \cos(k_y) - \mu_2
\label{2}
\end{equation}
representing the orthorhombic CuO chains.  In (\ref{1}) and (\ref{2}), 
$8\sigma_1$   and
$4\sigma_2$   are the band widths, $B$ is a next nearest neighbour hopping 
amplitude in
units $\sigma_1$, and $\mu_1$   and $\mu_2$ are chemical potentials
related to the fillings on the chains and planes.   This  model  is
already  quite  complicated  to  deal  with  and final results are
obtained  only  numerically.    Nevertheless,  it  needs   to   be
recognized  that  it  is  still  oversimplified  and may not apply
directly to YBCO.  For instance, in YBCO, there are  three  planes
per unit cell:  a bilayer of CuO$_2$  planes and a layer of CuO chains
in  between  them.   These 3 planes form the basic unit cell which
is  then  coupled  to the other unit cells through  additional,  
presumably   insulating,
layers.    Under  such  circumstances, estimates of the transverse
coupling $t_\perp$  provided from DC resistivity data may well reflect the
value of the intercell coupling (which could be incoherent) rather
than the plane-chain coupling within  the  cell  (which  could  be
coherent).    Treating  such complications is, however, beyond the
scope of this paper.  Here, we will assume only 2 planes but  take
one  to  be chains so as to introduce the observed orthorhombicity.
The transverse  coupling  $t_\perp$ therefore represents both the
chain-plane coupling and the intercell coupling.

The mean field Hamiltonian for our model is
\begin{equation}
H = \sum_{\bf k} C^\dagger_{\bf k} \, h({\bf k}) \, C_{\bf k}
\end{equation}
where $C^\dagger_{\bf k} = [c^\dagger_{1{\bf k}\uparrow},
c_{1-{\bf k}\downarrow},
c^\dagger_{2{\bf k}\uparrow},c_{2-{\bf k}\downarrow}]$ and $c^\dagger_{i{\bf k}\sigma}$
creates an electron of spin $\sigma$ and momentum ${\bf k}$ in the plane
($i=1$) or chain ($i=2$) layers.  The Hamiltonian matrix is
\begin{eqnarray}
h({\bf k}) = \left [
        \begin{array}{cccc}  \xi_1 & \Delta_{\bf k} & t({\bf k}) & 0 \\
                            \Delta_{\bf k} & -\xi_1 & 0 &-t({\bf k}) \\
		            t({\bf k}) & 0 &  \xi_2 & \Delta_{\bf k} \\
			    0 &-t({\bf k}) & \Delta_{\bf k} & -\xi_2 
	\end{array}
\right]
\label{9}
\end{eqnarray}
In Eq.~(\ref{9}) $t({\bf k}) = -2t_\perp \cos(k_z d/2)$, which follows from a 
tight-binding model of $c$-axis coupling.  The unit cell size is $d$ along the
$c$-axis.  The superconducting gap $\Delta_{\bf k}$ is chosen to be the same in
both the plane and chain layers.  We will not determine $\Delta_{\bf k}$
through solution of a gap equation but will instead be  guided  by
experiments  which  have  indicated that the gap on the planes and
chains are very similar. For our simple model we have
a  single $d$-wave gap characteristic of the entire system.  We make
no  attempt  to  relate  this  gap  to   the   basic   microscopic
interactions but simply postulate its existence.  

     In a previous paper, we derived a general expression for
the penetration depth of a system with two atoms per unit cell  (where
the two atoms form the chain and plane layers).\cite{v}  
The expression for the penetration depth required the two particle
Green's function:
\begin{eqnarray}
G_{\mu\nu}({\bf q},{\bf q}^\prime,\omega) &=& -\delta_{{\bf q},{\bf q}^\prime}
\frac{e^2}{\Omega}
\sum_{i,j,{\bf k}} [\hat{\gamma}_\mu({\bf k},{\bf k}+{\bf q})]_{ij}
[\hat{\gamma}_\mu({\bf k}+{\bf q},{\bf k})]_{ji} \nonumber \\
&&\times \frac{f[E_i({\bf k})]-f[E_j({\bf k}+{\bf q})]}{E_i({\bf k}) - \hbar \omega
-E_j({\bf k}+{\bf q})},
\label{3}
\end{eqnarray}
where $f(x)$ is the Fermi distribution function, 
$e$ is the electronic charge and the sum over $i$ and $j$ range from
1 to 4.  In (\ref{3}), the electromagnetic vertex 
$\hat{\gamma}_\mu({\bf k}+{\bf q},{\bf k})$ is a 
four  by four matrix related to the bare vertex function
${\gamma}_\mu({\bf k}+{\bf q},{\bf k})$ 
through the unitary transformation $U({\bf k})$  which  diagonalizes  the
Hamiltonian.  That is
\begin{equation}
\hat{\gamma}_\mu({\bf k},{\bf k}+{\bf q}) = U^\dagger({\bf k})   
{\gamma}_\mu({\bf k},{\bf k}+{\bf q})  U({\bf k}+{\bf q}).  
\label{4}
\end{equation}
The  quasiparticle energies $E_i({\bf k})$  are  the  eigenvalues of $h({\bf k})$:
$E_1=E_+$, $E_2=-E_+$, $E_3 = E_-$, $E_4=-E_-$, where
\begin{equation}
E_\pm = \sqrt{\epsilon_\pm^2 + \Delta_{\bf k}^2}
\end{equation}
and $\epsilon_\pm$ are the normal state band energies
\begin{equation}
\epsilon_\pm = \frac{\xi_1+\xi_2}{2} \pm \sqrt{\left[\frac{\xi_1-\xi_2}
{2}\right]^2 +t^2}.
\label{13}
\end{equation}

For a static external
electromagnetic vector potential $A({\bf r})$, the current is given by
\begin{equation}    
J_\mu({\bf r}) = - \sum_\nu  K_{\mu\nu} A_\nu ({\bf r})  
\label{5}
\end{equation}
where $(\mu,\nu)$ indicates spatial directions and  the  kernel $K_{\mu\nu}$
is given by
\begin{equation}
K_{\mu\nu} = \frac{1}{c} \left\{ \left. G_{\mu\nu}(0,0,0)
\right|_{\Delta=0} - G_{\mu\nu}(0,0,0) \right \}.
\label{6}
\end{equation}
In  equation  (\ref{6}), $c$ is the velocity of light, $\Omega$  
is the volume and
the first term in the curly bracket is  the  same  as  the  second
except  that  it is to be evaluated in the normal state with gap 
set equal to zero.  Note that only the ${\bf q}={\bf k}=0$  and  $\omega=0$  limit  has
been  taken  in  (\ref{6}).   This is sufficient for calculations of the
penetration depth which follows from 
\begin{equation}
  \frac{1}{\lambda_\mu^2(T)} = \frac{c}{4\pi}K_{\mu\mu}
\label{7}
\end{equation}

Our fundamental formula therefore, has the form
\begin{eqnarray}
G_{\mu\mu}(0,0,0) &=& -\frac{e^2}{\Omega}\sum_{i,j,{\bf k}} 
[\hat{\gamma}_\mu({\bf k},{\bf k})]_{ij}^2 \nonumber \\
&\times& \left [ \delta_{ij} \frac{\partial f(E_i)}{\partial E_i} -
[1-\delta_{ij}] \frac{f(E_i)-f(E_j)}{E_i-E_j} \right ].
\label{8}
\end{eqnarray}

The matrix $U({\bf k})$ which diagonalizes the Hamiltonian matrix can be
found using a two step process.
We  first introduce  a  unitary matrix $U_N$  which would diagonalize 
the normal state alone.  It has the form
\begin{eqnarray}
U_N = \left [
        \begin{array}{cccc}  u_n & 0 & v_n & 0 \\
                             0 & u_n & 0 & v_n \\
			     v_n & 0 & -u_n & 0\\
			     0 & v_n & 0 & -u_n
	\end{array}
\right]
\label{11}
\end{eqnarray}
with
\begin{equation}
u_n = \sqrt{\frac{\xi_1-\epsilon_-}{\epsilon_+-\epsilon_-}}
\end{equation}
and
\begin{equation}
v_n = \sqrt{\frac{\epsilon_+-\xi_1}{\epsilon_+-\epsilon_-}}.
\end{equation}
Then
\begin{eqnarray}
h^\prime &=& U^\dagger_N \, h \, U_N \nonumber \\
&=& \left [ 
        \begin{array}{cccc}  \epsilon_+ & \Delta_{\bf k} & 0 & 0 \\
			     \Delta_{\bf k} & -\epsilon_+& 0 & 0 \\
			     0 & 0 & \epsilon_- & \Delta_{\bf k} \\
			     0 & 0 & \Delta_{\bf k} & -\epsilon_-
	\end{array}
\right]
\label{14}
\end{eqnarray}

Matrix (\ref{14}) can finally be diagonalized by application of a second
unitary transformation $U_S$  to get
\begin{equation}
E_i = \left [U^\dagger_S \, h^\prime \, U_S \right ]_{ii}
\label{15}
\end{equation}
where
\begin{eqnarray}
U_S = \left [
        \begin{array}{cccc}  u_+ & v_+ & 0 & 0 \\
                             v_+ & -u_+ & 0 &0 \\
			     0 & 0 & u_- & v_- \\
			     0 & 0 & v_- &-u_-
	\end{array}
\right],
\label{16}
\end{eqnarray}
with
\begin{equation}
	u_\pm = \sqrt{\frac{E_\pm+\epsilon_\pm}{2E_\pm}}
\label{17}
\end{equation}
and
\begin{equation}
	v_\pm = \sqrt{\frac{E_\pm-\epsilon_\pm}{2E_\pm}}.
\label{18}
\end{equation}
where we stress again,  a  single  gap  has  been  introduced  for the
condensate on both the CuO$_2$ planes and CuO chains equally.  This
assumption was  justified  in  our  previous  work\cite{v}  and  has  
simplified  the mathematics enormously here.  

     There remains only to compute the transformed electromagnetic
vertex  $\hat{\gamma} = U^\dagger_S U^\dagger_N \gamma U_N U_S$
which  appears  in  equation  (\ref{8}).    The  bare
vertices are\cite{v}
\begin{equation} 
\gamma_x = \left[
	\begin{array}{cccc} v_{1x} & 0 & 0 & 0 \\
			    0 & v_{1x} & 0 & 0 \\
			    0 & 0 & 0 & 0 \\
			    0 & 0 & 0 & 0 
	\end{array}
\right],
\end{equation}
\begin{equation} 
\gamma_y = \left[
	\begin{array}{cccc} v_{1y} & 0 & 0 & 0 \\
			    0 & v_{1y} & 0 & 0 \\
			    0 & 0 & v_{2y} & 0 \\
			    0 & 0 & 0 & v_{2y} 
	\end{array}
\right],
\end{equation}
\begin{equation} 
\gamma_z = \left[
	\begin{array}{cccc} 0 & 0 & v_\perp & 0 \\
			    0 & 0 & 0 & v_\perp \\
			    v_\perp & 0 & 0 & 0 \\
			    0 & v_\perp & 0 & 0 
	\end{array}
\right],
\end{equation}
with
\begin{equation} 
	v_{i\mu} = \frac{1}{\hbar}\frac{\partial h_{ii}}{\partial k_\mu}
\end{equation}
and
\begin{equation} 	
	v_\perp =  \frac{1}{\hbar}\frac{\partial t}{\partial k_z}.
\end{equation}
It is straightforward to show that
\begin{equation} 
\hat{\gamma} = \left[
	\begin{array}{cccc} v_{\mu +} & 0 & \alpha v_\mu & -\beta v_\mu \\
			    0 & v_{\mu +} & \beta v_\mu & \alpha v_\mu \\
			    \alpha v_\mu & \beta v_\mu & v_{\mu-} & 0 \\
			    -\beta v_\mu & \alpha v_\mu & 0 & v_{\mu-} 
	\end{array}
\right],
\end{equation}
where
\begin{equation}
	v_{\mu\pm} = \frac{1}{\hbar}\frac{\partial \epsilon_\pm}
	{\partial k_\mu},
\end{equation}
and
\begin{mathletters}
\begin{equation}
	v_x = \frac{t}{\epsilon_+ - \epsilon_-} v_{1x},
\label{22a}
\end{equation}
\begin{equation}
	v_y = \frac{t}{\epsilon_+ - \epsilon_-} (v_{1y}+v_{2y}),
\label{22b}
\end{equation}
\begin{equation}
	v_z = \frac{\xi_2-\xi_1}{\epsilon_+ - \epsilon_-} v_\perp.
\label{22c}
\end{equation}
\end{mathletters}

The coherence factors $\alpha$ and $\beta$ have the simple form
\begin{mathletters}
\begin{equation}
	\alpha^2 = \frac{1}{2} \frac{E_+E_- + \epsilon_+\epsilon_-+\Delta^2}
	{E_+E_-}
\label{24a}
\end{equation}
\begin{equation}
	\beta^2 = \frac{1}{2} \frac{E_+E_- - \epsilon_+\epsilon_--\Delta^2}
	{E_+E_-}.
\label{24b}
\end{equation}
\end{mathletters}
Now that $\hat{\gamma}$ has been specified, $G_{\mu\mu}$ may be evaluated.

\section{Results and Discussion}
\label{three}
We  start  by rewriting equation (\ref{8}) (which gives the
penetration depth $\lambda(T)$ through (\ref{6}) and (\ref{7})) in the form
\begin{eqnarray}
G_{\mu\mu}(0,0,0) &=& -\frac{e^2}{\Omega}\sum_{\bf k} \left \{ \sum_i 
[\hat{\gamma}_\mu]^2_{ii}
\frac{\partial f(E_i)}{\partial E_i} \right .\nonumber \\
&&\left. -\sum_{i\neq j}[\hat{\gamma}_\mu]^2_{ij} \frac{f(E_i) - f(E_j)}
{E_i-E_j} \right \}
\label{25}
\end{eqnarray}
The first term in (\ref{25}) is the very familiar intraband contribution
and can be rewritten to read
\begin{equation}
-2\frac{e^2}{\Omega} \sum_{\bf k} \left\{ \left[ \frac{\partial \epsilon_+}{\partial k_\mu}
\right ]^2 \frac{\partial f(E_+)}{\partial E_+} -
\left[ \frac{\partial \epsilon_-}{\partial k_\mu}
\right ]^2 \frac{\partial f(E_-)}{\partial E_-} \right \}
\label{26}
\end{equation}
which is just the usual expression for the penetration depth in a
2 band model with energies given by equation (\ref{13}). For 
$t_\perp =0$ equation (\ref{26}) will make no contribution to $\lambda_z$ 
and we recover the
result for 2 decoupled bands. It is to be noted that the second
term in (\ref{25}), which is new and which gives the interband
transitions, will also give no contribution to $\lambda_z$  
when $t_\perp =0$ as it
must. This can be seen by noting that each of the vertex
elements $[\hat{\gamma}_\mu]_{ij}$, with $i\neq j$, is proportional to
$t_\perp$.
   The interband contribution to coming from the second term
in (\ref{25}) can be written as
\begin{eqnarray}
&& 4\frac{e^2}{\Omega}\sum_{\bf k} v_\mu^2 \left\{ \alpha^2 
\frac{f(E_+)-f(E_-)}
{E_+-E_-} \right . \nonumber \\
&& \left . \qquad - \beta^2 \frac{1-f(E_+)-f(E_-)}{E_++E_-} \right \},
\label{27}
\end{eqnarray}
where $\alpha$ and $\beta$ are given by Eqs.~(\ref{24a}) and (\ref{24b}).
These terms describe interband transitions.  The term with
$[f(E_+)-f(E_-)]/(E_+ -E_- )$  vanishes  at  $T=0$  and increases with
increasing $T$ while the one with $[1-f(E_+)-f(E_-)]/(E_+ +E_- )$ is 1 at
$T=0$ and decreases with increasing $T$. This competition between
these two terms leads to a much flatter temperature dependence for
$\lambda^{-2}(T)$ than does the more common intraband term (\ref{26}).  
Choosing
the $z$ direction, we note that the intraband part (\ref{26}) goes like $t^4$
because $v_z$ in (\ref{22c}) goes like $t_\perp^2$ while the interband
contribution (\ref{27}) goes only like $t_\perp^2$.  For sufficiently small
values of $t_\perp$, it is clear that the interband contribution will
dominate the $c$-axis penetration depth and that our intuition
based only on the intraband band term will no longer hold. Our
numerical work, to be described below, will tell us the value of
$t_\perp$ where this new regime of behaviour is reached.  Before
presenting results it is interesting to note that the in-plane
situation is quite different with regards to the dominance, or
lack there of, of the interband processes. This can be seen from
equation (\ref{26}) and the realization that, for the x or  y
directions, the intraband contribution is, to leading order,
independent of $t_\perp$, while from
(\ref{27}), we see that the interband part is proportional to 
$t_\perp^2$. Thus,
for small $t_\perp$, the interband contribution to the 
in-plane
penetration  depth will not be important compared with the
intraband contribution and the familiar behaviour will result. On
the other hand, when $t_\perp$ gets large, the interband contribution to
the in-plane penetration depth can become of some importance.

\begin{figure}[tb]
\begin{center}
\leavevmode
\epsfxsize 0.9\columnwidth
\epsffile{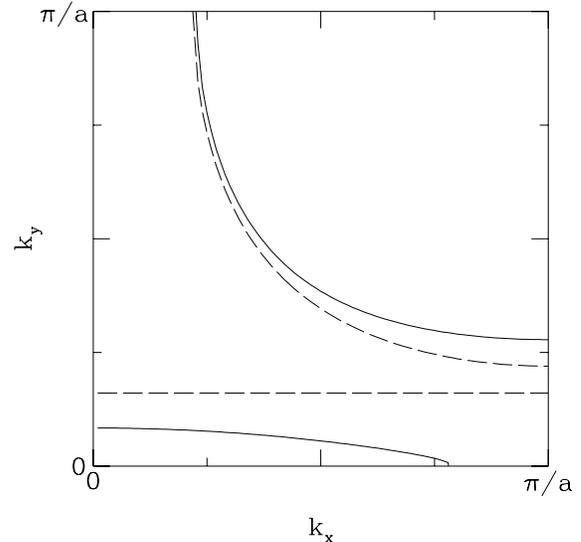}
\caption{Fermi surface for the chain-plane model. The parameters are
$\{\sigma_1,\sigma_2,\mu_1,\mu_2,t_\perp\} = \{70,100,-65,-175,20\} 
\mbox{ meV}$ and $B = 0.45$. The
dashed lines apply to the unperturbed case $t_\perp =0$ and represent
uncoupled plane and chain layers. They also describe the case
$k_z = \pi/d$ in equation (\protect\ref{4}), 
i.e. the Brillouin zone boundary. The
solid curves are for $k_z =0$ in (\protect\ref{4}), 
i.e. the central plane in the
B.Z. while the area between solid and dashed curves represent the
dispersion in the z-direction.}
\label{f1}
\end{center}
\end{figure}

In Fig.~\ref{f1}, we show our model band structure. We have taken
         $\{\sigma_1,\sigma_2,\mu_1,\mu_2,t_\perp\} 
= \{70,100,-65,-175,20\}$ meV with $B=0.45$.   The  dashed
line  gives  the  plane  and  chain  Fermi  surface contour in the
2-dimensional CuO$_2$  plane Brillouin zone as a function of  momentum
components $k_x$  and $k_y$. When the transverse matrix element $t_\perp$ is
taken into account, the Fermi surface acquires dispersion in the
$z$-direction.    When
$k_z$ is zero, $t(k_z)=2t$  and we get the hybridized
Fermi surface shown as the solid contours in Fig.~\ref{f1}.  On the other
hand, for $k_z$  at the zone edge ($k_z = \pi/d$), $t(k_z)=0$  
and  we  recover
the  dashed  contours.    For  other values of $k_z$, we obtain Fermi
surface contours intermediate between solid and dashed  curve  and
the  area  between  these 2 contours gives the $z$ dispersion of our
model.  This simplified Fermi surface will be kept  fixed  in  our
calculations and we present results for the penetration depth as a
function  of  temperature for all three directions $x,y,z$ 
(or, equivalently $a,b,c$).

\begin{figure}[tb]
\begin{center}
\leavevmode
\epsfxsize 0.9\columnwidth
\epsffile{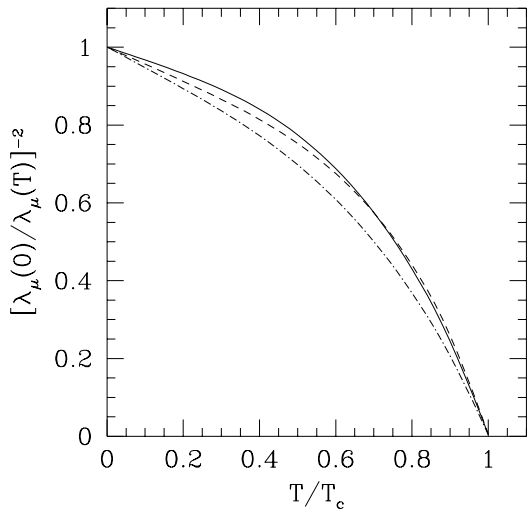}
\caption{
The inverse square penetration depth normalized to its value at $T=0$ 
in the a (dashed curve), b (dot-dashed curve) and c (solid curve) 
directions as a function of reduced
temperature $T/T_c$.  The transverse matrix element coupling the chain
and plane layers is $t_\perp =20$ meV.}
\label{f2}
\end{center}
\end{figure}

\begin{figure}[tb]
\begin{center}
\leavevmode
\epsfxsize 0.9\columnwidth
\epsffile{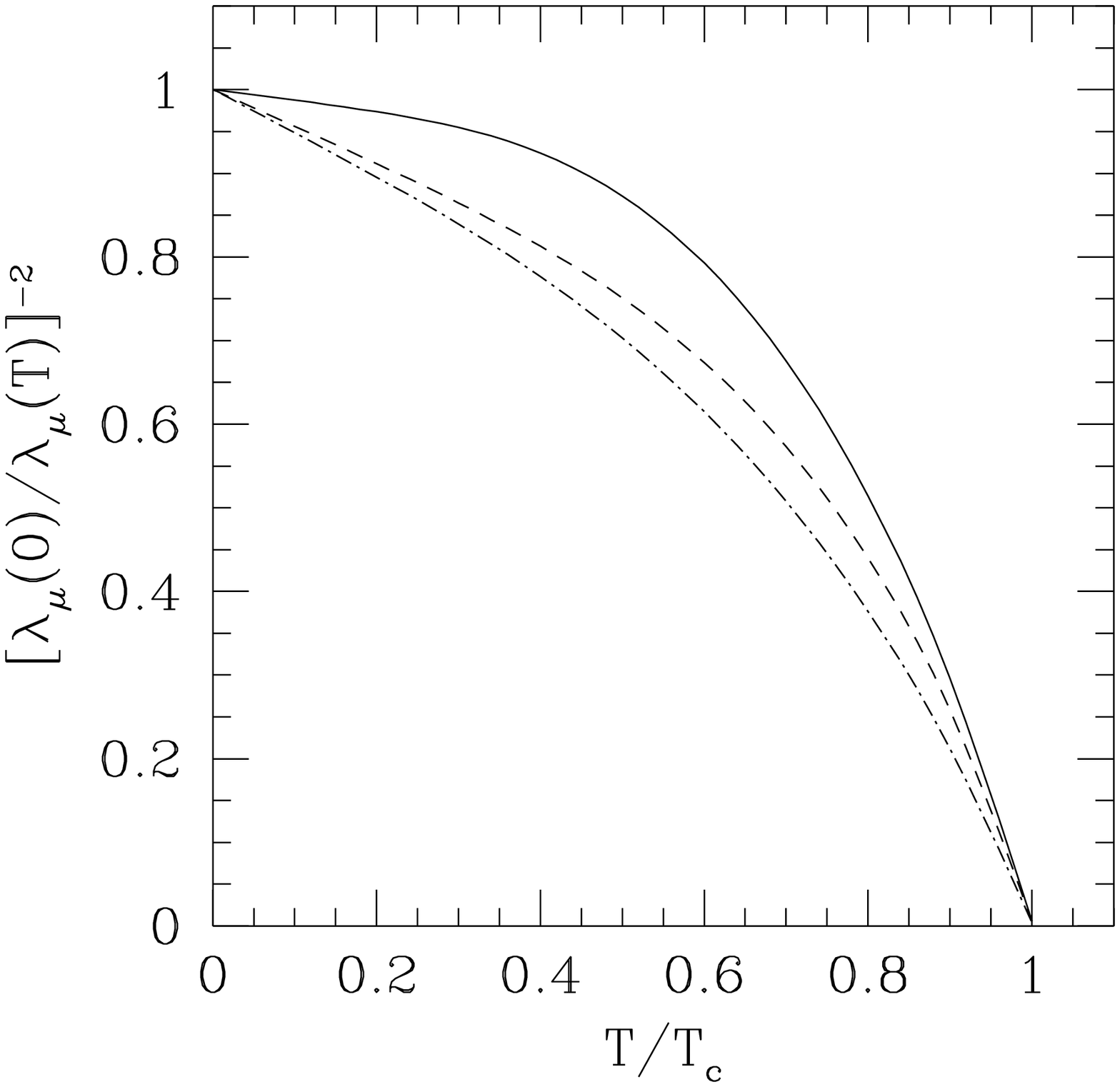}
\caption{As in Fig.~\protect\ref{f2}, penetration depth in the a 
(dashed curve), b (dot-dashed curve) and c (solid curve) 
directions but with $t_\perp =10$ meV.}
\label{f3}
\end{center}
\end{figure}

\begin{figure}[tb]
\begin{center}
\leavevmode
\epsfxsize 0.9\columnwidth
\epsffile{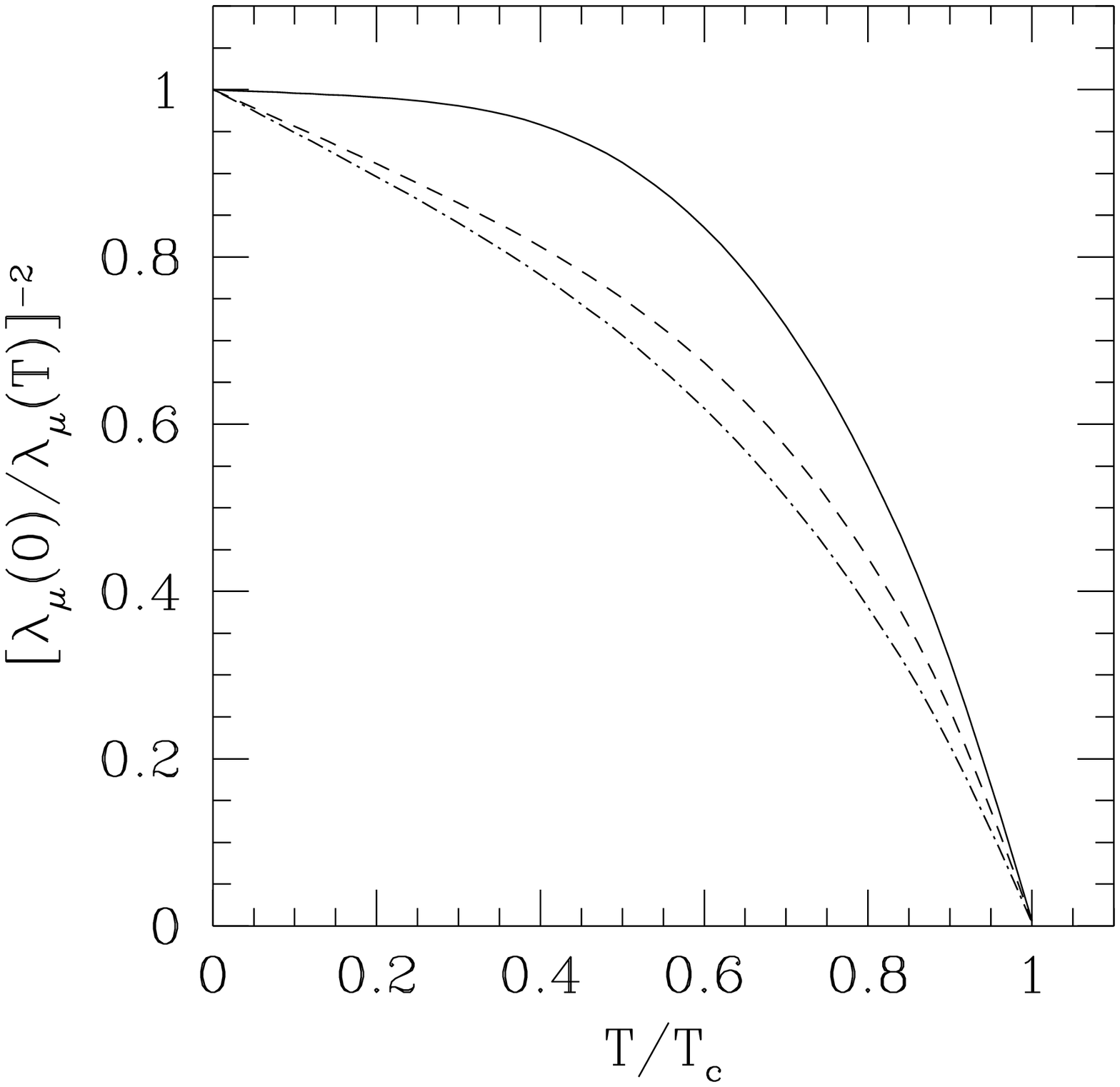}
\caption{As in Fig.~\protect\ref{f2}, penetration depth in the a 
(dashed curve), b (dot-dashed curve) and c (solid curve) 
directions but with $t_\perp =5$ meV.}
\label{f4}
\end{center}
\end{figure}

What is given in Figs.~\ref{f2}, \ref{f3} and \ref{f4}  are  the  
temperature  dependences  of
$[\lambda_\mu(0)/\lambda_\mu(T)]^2$ for $t_\perp =20$ meV, 10 meV and 5 meV, 
respectively.  In
all cases, we have used a $d$-wave gap of the form
\begin{equation}
   \Delta_{\bf k} = \Delta_0[\cos(k_x) - \cos(k_y)]
\label{28}
\end{equation}
over the 2-dimensional Brillouin zone  with $\Delta_0(T)$  taken  to
have   a   BCS   temperature   variation   but   scaled   so  that
$2\Delta_0(T=0)/k_B T_c =7$.  This value is chosen to agree  with  experiments
on  YBCO  which  find a rather large value for the gap to critical
temperature ratio.  This is also necessary in order to get  an  in-%
plane temperature dependence for the penetration depth which is in
reasonable  accord with experiment.  Before describing our results
and commenting on them, we stress that in YBCO$_x$ as a  function  of
oxygen  doping,  the Fermi surface of the chains will certainly be
changed in important ways because oxygen depletion occurs  on  the
chains as we go from the overdoped case $x=7$ to the underdoped case
with $x=6.6$.  At the same time, the number of holes on the planes as
well  as  on  the  chains will change so the value of the chemical
potential and displacement between bands, $\mu_1$ and $\mu_2$ will change as
will the filling factor in each band.  Such details  are  not  yet
known  and  in  the absence of such information, we have chosen to
keep the bands fixed as well as the filling which was  not  chosen
to  specifically model the overdoped or underdoped case.  Here, we
are primarily interested in understanding the effect of  interband
transitions  on  the  penetration  depth  and their variation with
chain-plane coupling $t_\perp$ which may change significantly with doping
although again this is not well known.

In Figs.~\ref{f2} to \ref{f4}, the dashed curve is for the $a$-direction, the
dot-dashed curve is for the $b$-direction and the solid for the
$c$-direction.  In the first case, $t_\perp =20$ meV and all three
directions show about the same temperature dependence  with
$\lambda_b^{-2}$ having the steepest slope at low $T$, $\lambda_a^{-2}$
having a slightly flatter slope and $\lambda_c^{-2}$ having the flattest 
slope.  This order is as observed in experiment. As the
transverse hopping matrix element $t_\perp$ is reduced in size, 
we see
that the in plane behaviour is not strongly affected but the
$c$-axis is.  It becomes flatter at low temperature and begins to
look much more like a usual $s$-wave case. This different behaviour
results when the interband terms (\ref{27}) start becoming dominant 
over
the more conventional intraband term (\ref{26}). We might expect this
to occur in the underdoped situation in YBCO where $t_\perp$ is expected
to be of order 5 meV or so. For optimally doped YBCO, interband
terms contribute less to the $c$-axis behaviour and the system
becomes more three dimensional.

\begin{figure}[tb]
\begin{center}
\leavevmode
\epsfxsize 0.9\columnwidth
\epsffile{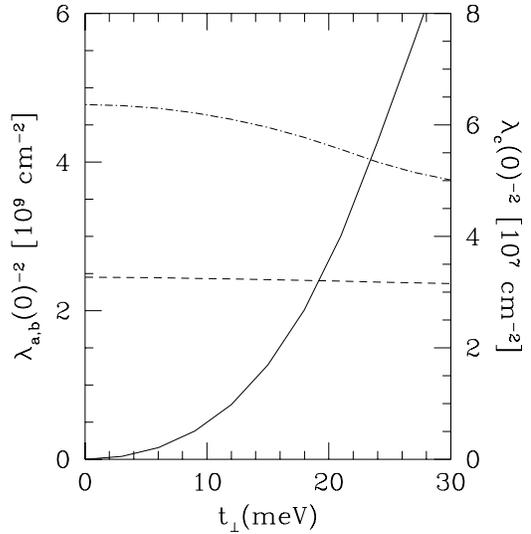}
\caption{Magnitude of the inverse penetration depth at $T=0$ as a
function of the plane-chain coupling.  The three curves are $\lambda_a(0)$
(dashed), $\lambda_b(0)$ (dot-dashed) and $\lambda_c(0)$ (solid).}
\label{f5}
\end{center}
\end{figure}

   In Fig.~\ref{f5}, we show the absolute value of the inverse square
of the zero temperature penetration depth $\lambda(0)^{-2}$ for the three
directions. Note the difference in scale between $\lambda_c^{-2}$ and 
$\lambda_{ab}^{-2}$. 
In the plot of $\lambda_c(0)$, we can see very clearly 
the $t_\perp^2$ dependence discussed earlier.
The magnitude of $\lambda_a^{-2}$ is smaller than
$\lambda_{b}^{-2}$, roughly by a factor of two because of the additional
screening current carried by the chains.   The
difference is reduced as $t_\perp$ increases. This is attributed to the
increase significance of the interband contribution to the
$b$-direction penetration depth as $t_\perp$ is increased. In the x
direction, the intraband term in (\ref{26}) goes like
\[
v_{1x}^2 \left [1+O\left( \frac{t^2}{(\xi_1-\xi_2)^2} \right ) \right ]
\]
for most of the Brillouin zone and the second term is small as is
the interband term (\ref{27}) which also varies as 
$O(t^2/(\xi_1-\xi_2)^2)$.
In the region where the plane and chain Fermi surface crosses this
argument does not apply, but $v_{1x}^2$ is small there and so we can
conclude that the terms in $t_\perp$ are not important. This does not
hold, however, for the y-direction because in that case $v_{1y}$ and
$v_{2y}$ are not necessarily small in regions of the Brillouin zone
where the corrections in $t_\perp$ are significant and we expect some
contribution to $\lambda^{-2}_{y}(T)$. 

\section{Conclusion}
\label{four}
   We have considered a model of two conducting planes per unit
cell, one with tetragonal symmetry representing a CuO$_2$ plane, 
the other with
orthorhombic symmetry, representing a CuO chain. 
A transverse hopping matrix element $t_\perp$ is
included which couples the two bands coherently and introduces the
third dimension. For such a model, new transitions enter the
electromagnetic properties that would not be present in the case
of  one  atom  per  unit  cell.   In  addition  to  the
usual intraband terms, there are terms which now correspond to
interband transitions.   We  have
investigated the effect of these terms on the value of the zero
temperature penetration depth and on its temperature dependence.
It is found that for the $a$-direction, the effects of $t_\perp$
are small while for the $b$-direction, along the chains, there can
be significant corrections to the intraband terms, although to
leading order $\lambda_b$ is still independent of $t_\perp$.
In this case, the corrections for the
interband terms go like $t_\perp^2$.

   For  the  $c$-direction (perpendicular to the planes), the
situation is quite different. The conventional intraband terms go
like $t_\perp^4$ while the interband ones go like $t_\perp^2$ so, 
for sufficiently
small values of $t_\perp$, the interband contribution eventually must
take over. We find that these have a new temperature dependence
quite  distinct from the linear law expected for a $d$-wave
superconductor at low temperature. They give a much flatter curve
at low temperature for the $c$-axis superfluid density.  Numerical
calculations indicate that the interband term will be important
when $t_\perp$ is of order 5-10 meV.  This value is reasonable for
underdoped YBCO.  In our previous work on the normal state, we found
that a pseudogap develops in the $c$ axis optical conductivity for such
small values of $t_\perp$.
This offers an alternate interpretation of the
$c$-axis data in the underdoped case, which does not rely on any
incoherent elastic scattering or $c$-axis out-of-plane phonon
assisted process.  Our calculations naturally give a linear
dependence in $a$- and $b$- directions and a much flatter
dependence for the $c$-axis as is observed. On the other hand, as
the coupling between the two planes is increased,
the system becomes much more 3 dimensional and the
temperature dependence of all three penetration depths become 
similar and are linear at low T for a $d$-wave gap. These findings
are all in general qualitative agreement with the data in YBCO.

As discussed in the text, however, our results can probably not be
applied to the oxides without modifications because our electronic
structure model is too simple. For example, the unit cell in YBCO
contains a CuO$_2$  bilayer as well as CuO chains. The coupling
between planes within a cell is likely to be quite different from
that between two cells while in our simplified model a single $t_\perp$
enters. Still, the model allows us to introduce and understand
better    the   role
that interband transitions can play in the $c$-axis penetration
depth. It exhibits many of the qualitative properties observed in
YBCO.

\section*{Acknowledgements}
   Research supported in part by the Canadian Institute for
Advanced Research (CIAR) and by the Natural  Sciences  and
Engineering Research Council of Canada (NSERC).

\end{document}